\documentclass[aps,prl,twocolumn,superscriptaddress,showpacs]{revtex4}
\usepackage{graphicx}

\begin{document}


\title{Short-range magnetic ordering process for the triangular lattice 
compound NiGa$_2$S$_4$: a positive muon spin rotation and relaxation study}



\author {A.~Yaouanc}
\affiliation{CEA/DSM/Institut Nanosciences et Cryog\'enie, 38054 Grenoble, France}

\author{P.~Dalmas de R\'eotier}
\affiliation{CEA/DSM/Institut Nanosciences et Cryog\'enie, 38054 Grenoble, France}

\author{Y. Chapuis }
\affiliation{CEA/DSM/Institut Nanosciences et Cryog\'enie, 38054 Grenoble, France}

\author{C. Marin}
\affiliation{CEA/DSM/Institut Nanosciences et Cryog\'enie, 38054 Grenoble, France}

\author{G. Lapertot}
\affiliation{CEA/DSM/Institut Nanosciences et Cryog\'enie, 38054 Grenoble, France}

\author{A.~Cervellino}
\altaffiliation[On leave from]{
CNR, Istituto di Cristallografia (CNR-IC), 70126 Bari, Italy.}
\affiliation{Laboratory for Neutron Scattering, 
ETH Z\"urich and Paul Scherrer Institute, 5232 Villigen-PSI, Switzerland}

\author{A.~Amato}
\affiliation{Laboratory for Muon-Spin Spectroscopy, 
Paul Scherrer Institute, 5232 Villigen-PSI, Switzerland}



\date{\today}

\begin{abstract}

We report a study of the triangular lattice Heisenberg magnet NiGa$_2$S$_4$ 
by the 
positive muon spin rotation and relaxation techniques. We unravel three 
temperature regimes: (i) 
below $T_{\rm c} = 9.2 \, (2)$ K a spontaneous static magnetic field at the 
muon site is observed and the spin dynamics is appreciable: the time scale of 
the modes we probe is $\simeq$ 7~ns; (ii) an unconventional stretched 
exponential relaxation function is found for $T_{\rm c} < T < T_{\rm cross}$ 
where $T_{\rm cross} = 12.6$~K, which is a signature of 
a multichannel relaxation for this temperature range; (iii) 
above $T_{\rm cross}$, the relaxation is exponential as expected for a 
conventional
compound. The transition at $T_{\rm c}$ is of the continuous type. 
It occurs at a temperature slightly smaller than the temperature at which the 
specific heat displays a maximum at low temperature. This is reminiscent
of the behavior expected for the Berezinskii-Kosterlitz-Thouless
transition. We argue that these results reflect the presence of topological 
defects above $T_{\rm c}$. 

\end{abstract}


\maketitle

On cooling, in the same way as liquids, magnetic materials, usually crystallize to form long-range 
periodic arrays. 
However, magnetic materials with antiferromagnetically coupled spins located 
on triangular motifs 
exhibit geometrical magnetic frustration which may prevent the crystallization to occur
\cite{Ramirez01}. Such materials are a fertile ground for the emergence of novel spin-disordered states such 
as spin liquid or spin glass even without crystalline disorder. The simplest example of geometrical 
frustration, stacked two-dimensional triangular lattices with a single 
magnetic ion per unit cell, have been studied intensively \cite{Collings97}.  
NiGa$_2$S$_4$ is a rare example of such a two-dimensional antiferromagnet
which does not exhibit a long-range magnetic order at low temperature and
which is characterized by gapless excitations
\cite{Nakatsuji05,Nakatsuji07a}. These physical properties lead naturally to the assumption 
that its ground state is a spin liquid. Here we report muon spin rotation and relaxation measurements ($\mu$SR) which 
show that a spontaneous static magnetic field appears below $T_{\rm c} = 9.2 \, (2)$~K where
an appreciable spin dynamics is measured. In addition, we find the spin 
dynamics to be unconventional in the
temperature range $T_{\rm c} < T < T_{\rm cross}$ where 
$T_{\rm cross} = 12.6$~K.

Polycrystalline powder of NiGa$_2$S$_4$ has been obtained from a solid-vapor reaction using a 
stoichiometric mixture of pure elements.
The synthesis took place 
in an evacuated silica ampoule. A slow heat treatment over several days up to 1000$^\circ$C 
has been performed in respect to the high sulphur vapor pressure \cite{Lutz86}. After a final 
grinding, the powder was treated for a week at 1000$^\circ$C.

NiGa$_2$S$_4$  is a chalcogenide magnetic insulator with the Ni$^{2+}$ 
magnetic ions (spin $S = 1$) sitting on 
an regular triangular lattice. The interactions are of the Heisenberg type, 
referring to the isotropic Curie constant \cite{Nakatsuji07a}. 
The crystal
structure consists of two GaS layers and a central NiS$_2$ layer stacked along the $c$-axis. A Rietveld 
refinement of a neutron powder diffraction pattern
recorded at 50~K at the cold neutron powder 
diffractometer DMC of the SINQ facility at the Paul Scherrer Institute (Villigen, Switzerland)
is consistent with the $P{\bar 3}m1$ space group. The 
lattice parameters are $a$ = 0.3619~nm and $c$ = 1.1967~nm,
in agreement with previous results \cite{Nakatsuji05}. 
The refinement is consistent with the nominal stoichiometry and no impurity phase is detected (detection limit: $1 \, \%$). 

Further characterizations of our sample have been done by zero-field 
specific heat and susceptibility measurements.  
The magnetic specific heat $C_{\rm m}$ (divided by the temperature)
is displayed in Fig.~\ref{data_basic}. It is
in very reasonable agreement with the data of Nakatsuji and coworkers \cite{Nakatsuji05}. 
As these authors reported, we also find that the susceptibility recorded 
under a field of $0.01$~T displays
a weak kink at $T_\chi \simeq 8$~K. 
  
\begin{figure}
\includegraphics[scale=0.8]{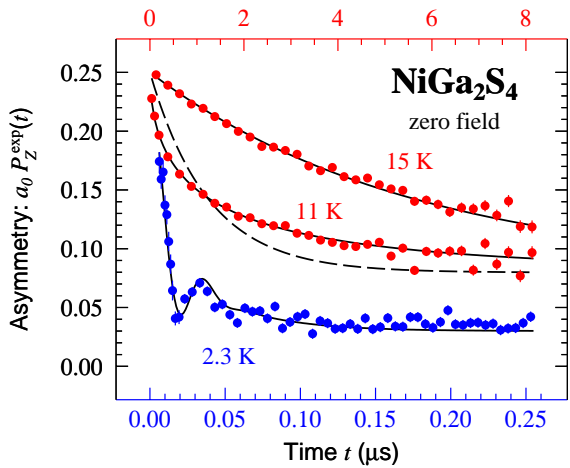}
\caption{(color online).
Three zero-field $\mu$SR spectra recorded for a powder sample of
NiGa$_2$S$_4$ with the GPS spectrometer
of the S$\mu$S facility at the Paul Scherrer Institute, Villigen, Switzerland.
The spectra probe three distinct temperature regimes.
At the bottom we display a spectrum recorded deep into the 
short-range ordered state. The solid line is the result of a fit to a function
given in the main text. 
The spectrum in the middle has been taken at 11~K where the fit relaxation 
function is a stretched exponential (full line). 
For reference, the dashed line displays the best fit to an exponential 
relaxation. Such an exponential function is observed at 15~K (top spectrum).
For clarity, the two high temperature
spectra are shifted up by $0.05$ relative to the third one. 
Their time scale is given by the top horizontal axis, 
while the scale for the low temperature spectrum
is at the bottom horizontal axis: the range spanned by these two time
scales differs by a factor $\sim 30$.}
\label{muon_spectra}
\end{figure}

\begin{figure}
\includegraphics[scale=0.8]{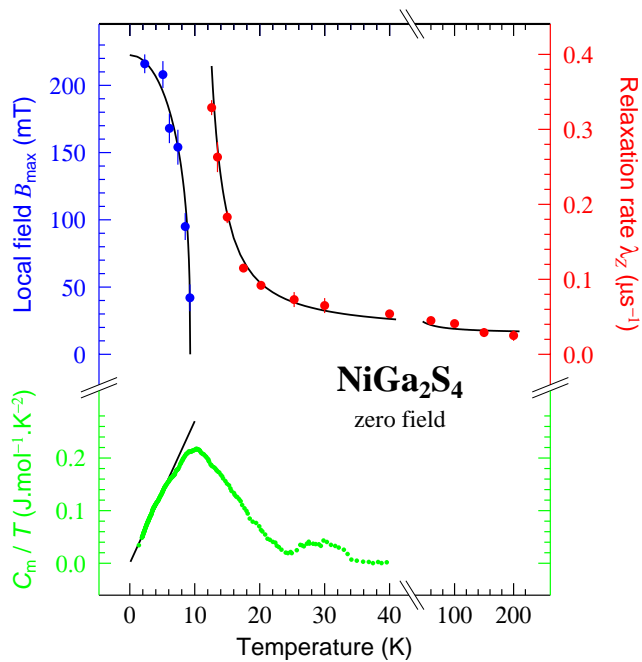}
\caption{(color online). Upper panel: thermal dependence of two parameters 
extracted from $\mu$SR spectra. On the left-hand side is presented the maximum 
of the static local magnetic field at the muon site, $B_{\rm max}$, 
for $T < T_{\rm stat}$.  On the right-hand side of the same panel is 
displayed the relaxation rate, $\lambda_Z$, 
for $T>T_{\rm cross}$, {\sl i.e.{}} 
when the relaxation function is exponential. 
The full lines are results of fits described in the main text.
Lower panel: zero-field magnetic specific heat 
divided by the temperature, $C_{\rm m}/T$, deduced from the measured 
specific heat using a dynamic adiabatic technique after subtraction of the lattice 
contribution. The full line is obtained for a linear dependence of $C_{\rm m}/T$.
}
\label{data_basic}
\end{figure}

Now we report on our zero-field $\mu$SR data; see Refs.~\cite{Dalmas97,Dalmas04} for an 
introduction to the $\mu$SR techniques. A spectrum is expressed as $a_{\rm 0} P^{\rm exp}_Z(t)$ 
where $a_{\rm 0}$ is an amplitude or asymmetry and $P^{\rm exp}_Z(t)$ a polarization function. 
In  Fig.~\ref{muon_spectra} we display three spectra which probe the three temperature regimes 
that we have unveiled. 

At variance with Ref.~[\onlinecite{MacLaughlin07}], a strongly damped oscillation is observed 
at low temperature where nano-scale magnetic correlations have been detected by neutron 
diffraction \cite{Nakatsuji05}; see the spectrum at the bottom of Fig.~\ref{muon_spectra}. 
Remarkably, this oscillation which reflects the presence of a spontaneous 
internal field, vanishes at a temperature which is about half the 
temperature at which the neutron magnetic reflections disappear. This is further discussed
below.
Our observation is also technically interesting. It shows that, contrary to common wisdom, the 
detection for a material of a zero-field $\mu$SR oscillation is not a 
fingerprint of a long-range magnetic order.

The spectrum at 2.3 K has a steep slope for time $t < 0.02 \, \mu {\rm s}$. 
Such a shape is typical for an incommensurately ordered
magnet which shows a characteristic field distribution at the muon 
site \cite{Overhauser60}. This translates in time-space to a Bessel 
function rather than cosine oscillations \cite{Dalmas97}. 
Our observation is not surprising since it has been established by neutron 
diffraction diffraction that the magnetic
structure is indeed incommensurate \cite{Nakatsuji05}. The spectrum is
described by the sum of two components for the compound and a 
third component which accounts for the muons missing the sample and stopped
in its surroundings:  
$a_{\rm 0} P^{\rm exp}_Z(t)= a_{\rm os} J_0(\gamma_\mu B_{\rm max}t )
\exp(-\gamma_\mu^2\Delta^2t^2/2) + a_{\rm rel} \exp(-\lambda_Z t) + 
a_{\rm bg}$. $J_0$ is the zeroth-order Bessel function of the first kind,
$B_{\rm max}$ stands for the maximum of the spontaneous static local magnetic 
field distribution at the muon site, $\Delta^2$ characterizes the broadening of the 
probed field distribution and $\gamma_\mu$ is the muon gyromagnetic ratio 
($\gamma_\mu$ = 851.615 Mrad s$^{-1}$ T$^{-1}$). The component of
amplitude $a_{\rm rel}$ gauges the relaxation of the muons sensing a
field parallel to their initial polarization. The associated
spin-lattice relaxation rate is denoted $\lambda_Z$. The asymmetry ratio of 
the first two 
components is $a_{\rm os}/a_{\rm rel} \simeq 1.6\, (2)$ to be compared to
an expected value of 2. A weak texturation of the sample could explain
this small deviation. 

Figure \ref{data_basic} shows $B_{\rm max}(T)$.
The full line is a fit to the phenomenological function 
$B_{\rm max}(T) = B_{\rm max} (0) [1 -(T/T_{\rm stat})^b]^\beta$ 
with $b=2$ and $\beta = 0.365$. We find $B_{\rm max} (0) = 222 \, (20)$ mT
and the temperature for which the spontaneous field vanishes 
$T_{\rm stat} = 9.3 \, (1)$ K. 
The magnitude of $B_{\rm max}$ reflecting the ordered magnetic moment, the 
transition at $T_{\rm stat}$ is consistent with a continuous 
transition. The field distribution is quite broadened,
$\Delta/B_{\rm max} \simeq 0.15$, in accord with our observation of only
two oscillations.
The detection of these oscillation provides a bound for the time scale of the 
magnetic 
correlations \cite{Dalmas06}: $ \tau_c \gtrsim 1/(\gamma_\mu B_{\rm max})$ = 
5~ns.   
Last but not least, $\lambda_Z$ is far from being negligible. Its temperature dependence mimics
$B_{\rm max}(T) $ with a value of $\sim$ 12~$\mu$s$^{-1}$ at low temperature. 
Because of the finite $\lambda_Z$ value, we can derive more than a bound for
$\tau_c$. From the relation $\lambda_Z = 2 \gamma_\mu^2 \Delta^2 \tau_c$ 
\cite{Dalmas97}, we compute 
$\tau_c \simeq 7$~ns, since $\Delta = 35 \, (4)$~mT. This value and the 
previous bound are consistent. Therefore 
the time scale of the magnetic correlations is $\tau_c \simeq$ 7~ns at low
temperature. 
The neutron scattering results, which set a lower bound of $\tau_{\rm min}$ 
= 0.3~ns for this time, are fully consistent \cite{Nakatsuji05}.

The spectrum in the middle of Fig.~\ref{muon_spectra} has been recorded at a 
temperature above $T_{\rm stat}$ and in a regime where the relaxation is 
stretched exponential-like:
$a_{\rm 0} P^{\rm exp}_Z(t)= a_{\rm rel} \exp[-(\lambda_Z t)^{\beta_\mu}] + 
a_{\rm bg}$. This occurs in the range $T_{\rm stat}< T< T_{\rm cross} =12.6$~K for which
three spectra have been recorded: at 10, 11 and 11.6~K. We get 
consistently $\beta_\mu = 0.56 \, (3)$.

The third spectrum in Fig.~\ref{muon_spectra} corresponds to the case where 
the sample is magnetically homogeneous and 
the magnetic fluctuations are fast enough to give rise to an exponential 
relaxation function {\sl i.e.{}} $\beta_\mu$ = 1. As seen in 
Fig.~\ref{data_basic}, $\lambda_Z$ increases for decreasing temperature.
This is the signature of the slowing down of the magnetic fluctuations. 
The relaxation rate is fitted to the formula
$\lambda_Z$ = $\lambda_0 [T/(T-T_{\rm dyn})]^2$ 
where $\lambda_0 = 0.028 \, (3)$~$\mu$s$^{-1}$ and 
$T_{\rm dyn} = 9.2 \, (2)$~K. This 
formula is derived from the random phase approximation (RPA) 
\cite{Yaouanc_Dalmas_unpublished_1}.
It is expected to be valid for $T > T_{\rm stat}$ provided $T$ is not too
close to $T_{\rm stat}$ which is the case for $T > T_{\rm cross}$.
As expected, the characteristic 
temperature $T_{\rm dyn}$ introduced in this model is equal 
to $T_{\rm stat}$ within error bars: we 
henceforth replace $T_{\rm dyn}$ and $T_{\rm stat}$ with a unique
parameter $T_{\rm c} = 9.2 \, (2)$ K.

We now discuss the implications of our results for the physics of 
NiGa$_2$S$_4$.

We note that $C_{\rm m}$ displays a rounded peak with a maximum at 
$\sim 13$~K, {\sl i.e.} just above $T_{\rm c}$.
Such a behavior is reminiscent of the Berezinskii-Kosterlitz-Thouless
transition temperature relevant for a planar XY magnet; see Fig. 9.4.3 of 
Ref.~[\onlinecite{Chaikin95}]. 
This is a key result from our study.

The relaxation function from a magnetically homogeneous sample is expected 
to be exponential when measured in its paramagnetic state; see e.g. elemental 
Ni \cite{Nishiyama84} or Fe \cite{Herlach86} or the intermetallics 
GdNi$_5$ \cite{Yaouanc96}. A stretched exponential 
relaxation has been observed for a wide variety of physical quantities in many different 
systems and research areas \cite{Phillips96}. It arises from a 
continuous sum of exponential decays \cite{Johnston06}. A square-root relaxation of the 
Nuclear Magnetic Resonance \cite{McHenry72} and $\mu$SR \cite{Uemura80} relaxation function 
is also observed for spatially disordered systems. It stems, e.g. for spin-glass 
materials, from the distributed nature of the coupling of the probe to its environment. 
Our sample is spatially homogeneous. We ascribe the stretched exponential 
relaxation  
for $T_{\rm c} < T < T_{\rm cross}$  to a multichannel relaxation process.

Nakatsuji {\it et al} have presented the temperature dependence of the difference $\Delta I(T)$ 
between the elastic powder neutron scattering intensity recorded at $T$ and $50$~K, for 
wave vector $q_{\rm c}$ = 0.58~\AA$^{-1}$ 
\cite{Nakatsuji05}. $\Delta I(T)$ decreases smoothly up to $\sim$ 18~K where it becomes
negligible. This temperature is quite different from $T_{\rm c}$, where the 
muon spontaneous field vanishes, and from $T_\chi$. These differences 
are not surprising, given the different magnetic
modes which are probed by the three techniques. The neutron intensity 
integrates the $q_{\rm c}$-correlations for times longer than $\tau_{\rm min}$.
$B_{\rm max}$ is built up from the sum of the dipole magnetic fields at the 
muon site due to a restricted $q$ range of the Ni$^{2+}$ magnetic 
moment Fourier components, and characterized by a time scale longer by an 
order of magnitude than $\tau_{\rm min}$. d.c susceptibility probes an even
longer time scale. That 
$\Delta I(T)$ vanishes only at $\sim 18$~K suggests a relatively large 
continuous spectrum
of magnetic fluctuations in NiGa$_2$s$_4$. This could be a common property of geometrically
frustrated magnets  since it has already been encountered for 
Tb$_2$Sn$_2$O$_7$ \cite{Dalmas06,Chapuis07,Rule07}. Because of the notable reduction of the 
Ni$^{2+}$ ($S$ = 1)
magnetic moment as measured by neutron diffraction ($\sim 25 \%$), the spectral weight 
must also extend to time scales smaller than $\tau_{\rm min} $.

Previous thermodynamic and neutron results \cite{Nakatsuji05} and the present 
$\mu$SR measurements on NiGa$_2$S$_4$ demonstrate that this two-dimensional
Heisenberg triangular lattice antiferromagnet has unique low-temperature 
properties:
(i) while its Curie-Weiss temperature is rather 
large, $\Theta_{\rm CW} = -80 \, (2)$~K, it does not display a long-range 
magnetic 
ordering, but only an incommensurate short-range order with a nano-scale 
correlation 
length and a spontaneous static interstitial magnetic field below 
$T_{\rm c} = 9.2 \, (2)$~K; (ii) its ground state is highly degenerate; 
(iii) there is a slowing down of 
magnetic fluctuations as the compound is cooled down through $T_{\rm c}$ 
rather than a spin-freezing as observed for canonical spin glasses. 
In fact, we have found that NiGa$_2$S$_4$ exhibits a conventional 
paramagnetic spin dynamics down to $T_{\rm cross}$, where an effective 
multichannel relaxation 
process sets in down to $T_{\rm c}$. A finite spin dynamics is detected below 
$T_{\rm c}$.

Before discussing theoretical proposals for the ground state of an exact 
triangular lattice of isotropic spins in light of the experimental results, 
we note that different techniques show that the exchange interaction between
third neighbors 
is strong \cite{Mazin07,Takubo07}. This result provides a clue for the
existence of
strong frustration, in agreement with experimental data \cite{Nakatsuji05}. 

A hidden order parameter associated to an ordered nematic phase can be 
proposed for this system \cite{Chandra91}, on the ground of the large coherent 
length inferred from an analysis of the specific heat data \cite{Nakatsuji05}.
In this case there are no long-range two-spin correlations and only the 
quadrupole moments of the Ni$^{2+}$ ions order in a long-range 
manner. This phase is classified as a spin liquid. It can be stable on a 
two-dimensional triangular lattice and massless excitations are 
present \cite{Tsunetsugu06,Lauchli06,Bhattacharjee06,Li07a}.
However, at least for the model available, strong biquadratic interactions 
are required. 

A second possibility for the ground state relies on the presence of residual 
defects in the system, the effect of which is enhanced by frustration. This 
picture would naturally explain the stretched exponential muon
relaxation function for $T_{\rm c} < T < T_{\rm cross}$. 
However, the substitution of only $1 \%$ of Zn for Ni dramatically affects 
the specific heat \cite{Nakatsuji07a}. Hence, the amount
of residual defects is probably small. In addition, we are not aware of a 
model calculation which would explain the limited correlation length and the 
persistence of relatively fast fluctuation modes below $T_{\rm c}$.

A third candidate model attributes the unique properties of NiGa$_2$S$_4$ to 
topological defects inherent to Heisenberg triangular two-dimensional systems, the 
so-called  $Z_2$-vortices \cite{Kawamura84,Kawamura07}. Gapless 
excitation modes and a nearly constant susceptibility are predicted 
\cite{Fujimoto06}. 
Based on Monte Carlo simulations, a phase transition was suggested 
\cite{Kawamura84}. However, because of spin-wave interactions, 
the correlation length is finite 
\cite{Azaria92,Southern93,Wintel95,Fujimoto06}.
It is therefore tempting to attribute the transition at $T_{\rm c}$ to the 
dissociation of  
the $Z_2$-vortices and $T_{\rm cross}$ to the crossover temperature where the spin dynamics 
starts to be driven by usual Heisenberg  spin fluctuations. Remarkably, the $Z_2$-vortices
manifest themselves in the temperature vicinity where $C_{\rm m}$ has a rounded peak. 
The multichannel 
relaxation for $T_{\rm c}< T < T_{\rm cross}$ reflects the magnetic disorder 
induced by the unbinding of $Z_2$-vortices.

In conclusion, we have found that, on cooling, the frustrated two dimensional 
triangular lattice compound NiGa$_2$S$_4$ first behaves as a conventional 
magnetic compound ordering at $T_{\rm c} = 9.2 \, (2)$~K. 
This behavior is however observed only down to 12.6~K ({\sl i.e.{}} 
$\sim 3.4$~K above $T_{\rm c}$). Below
this temperature the $\mu$SR relaxation is stretched exponential-like. 
This result is interpreted as the
 signature of an intrinsic property of the triangular system such as the 
$Z_2$-vortices. The dynamics is never frozen, even far below $T_{\rm c}$.
Finally, the transition at $T_{\rm c}$ from the short range ordered to the 
paramagnetic phase is of the continuous type and
occurs at a temperature just below that of the specific heat 
bump, reminding the Berezinskii-Kosterlitz-Thouless transition.

For a further insight into the properties of NiGa$_2$S$_4$, it is necessary 
to examine the wavevector dependence of the fluctuating magnetic modes
below $T_{\rm c}$. 
On the theoretical front, an interesting result would be to determine whether 
modes with a temperature dependent gap vanishing at $T=0$~K are possible 
for the triangular lattice as it seems to be the case for the kagom\'e 
structure
\cite{Li07b}. This would provide an explanation for the observed persistent 
spin dynamics for $T \ll T_{\rm c}$ \cite{Yaouanc05a}.

{\em Note added}.
A related report including 
$\mu$SR data recorded on this material
is also available \cite{Takeya08}. We note that the time range available
at the $\mu$SR facility used to record 
these data does not allow
the authors to evidence the spontaneous muon spin precession that we have
observed.

\bibliography{reference}

\begin{thebibliography}{37}
\expandafter\ifx\csname natexlab\endcsname\relax\def\natexlab#1{#1}\fi
\expandafter\ifx\csname bibnamefont\endcsname\relax
  \def\bibnamefont#1{#1}\fi
\expandafter\ifx\csname bibfnamefont\endcsname\relax
  \def\bibfnamefont#1{#1}\fi
\expandafter\ifx\csname citenamefont\endcsname\relax
  \def\citenamefont#1{#1}\fi
\expandafter\ifx\csname url\endcsname\relax
  \def\url#1{\texttt{#1}}\fi
\expandafter\ifx\csname urlprefix\endcsname\relax\def\urlprefix{URL }\fi
\providecommand{\bibinfo}[2]{#2}
\providecommand{\eprint}[2][]{\url{#2}}

\bibitem[{\citenamefont{Ramirez}(2001)}]{Ramirez01}
\bibinfo{author}{\bibfnamefont{A.~P.} \bibnamefont{Ramirez}}, in
  \emph{\bibinfo{booktitle}{Handbook of Magnetic Materials}}, edited by
  \bibinfo{editor}{\bibfnamefont{K.~H.~J.} \bibnamefont{Buschow}}
  (\bibinfo{publisher}{Elsevier}, \bibinfo{year}{2001}),
  vol.~\bibinfo{volume}{13}.

\bibitem[{\citenamefont{Collings and Petrenko}(1997)}]{Collings97}
\bibinfo{author}{\bibfnamefont{M.~F.} \bibnamefont{Collings}} \bibnamefont{and}
  \bibinfo{author}{\bibfnamefont{O.~A.} \bibnamefont{Petrenko}},
  \bibinfo{journal}{Can. J. Phys.} \textbf{\bibinfo{volume}{75}},
  \bibinfo{pages}{605} (\bibinfo{year}{1997}).

\bibitem[{\citenamefont{Nakatsuji et~al.}(2005)\citenamefont{Nakatsuji, Nambu,
  Tonomura, Sakai, Jonas, Broholm, Tsunetsugu, Qiu, and Maeno}}]{Nakatsuji05}
\bibinfo{author}{\bibfnamefont{S.}~\bibnamefont{Nakatsuji}},
  \bibinfo{author}{\bibfnamefont{Y.}~\bibnamefont{Nambu}},
  \bibinfo{author}{\bibfnamefont{H.}~\bibnamefont{Tonomura}},
  \bibinfo{author}{\bibfnamefont{O.}~\bibnamefont{Sakai}},
  \bibinfo{author}{\bibfnamefont{S.}~\bibnamefont{Jonas}},
  \bibinfo{author}{\bibfnamefont{C.}~\bibnamefont{Broholm}},
  \bibinfo{author}{\bibfnamefont{H.}~\bibnamefont{Tsunetsugu}},
  \bibinfo{author}{\bibfnamefont{Y.}~\bibnamefont{Qiu}}, \bibnamefont{and}
  \bibinfo{author}{\bibfnamefont{Y.}~\bibnamefont{Maeno}},
  \bibinfo{journal}{Science} \textbf{\bibinfo{volume}{309}},
  \bibinfo{pages}{1697} (\bibinfo{year}{2005}).

\bibitem[{\citenamefont{Nakatsuji et~al.}(2007)\citenamefont{Nakatsuji, Nambu,
  Onuma, , Jonas, Broholm, and Maeno}}]{Nakatsuji07a}
\bibinfo{author}{\bibfnamefont{S.}~\bibnamefont{Nakatsuji}},
  \bibinfo{author}{\bibfnamefont{Y.}~\bibnamefont{Nambu}},
  \bibinfo{author}{\bibfnamefont{K.}~\bibnamefont{Onuma}}, ,
  \bibinfo{author}{\bibfnamefont{S.}~\bibnamefont{Jonas}},
  \bibinfo{author}{\bibfnamefont{C.}~\bibnamefont{Broholm}}, \bibnamefont{and}
  \bibinfo{author}{\bibfnamefont{Y.}~\bibnamefont{Maeno}}, \bibinfo{journal}{J.
  Phys.: Condens, Matter} \textbf{\bibinfo{volume}{19}},
  \bibinfo{pages}{145232} (\bibinfo{year}{2007}).

\bibitem[{\citenamefont{Lutz et~al.}(1986)\citenamefont{Lutz, Buchmeier, and
  Siwert}}]{Lutz86}
\bibinfo{author}{\bibfnamefont{H.~D.} \bibnamefont{Lutz}},
  \bibinfo{author}{\bibfnamefont{W.}~\bibnamefont{Buchmeier}},
  \bibnamefont{and} \bibinfo{author}{\bibfnamefont{H.}~\bibnamefont{Siwert}},
  \bibinfo{journal}{Z. Anorg. Allg. Chem.} \textbf{\bibinfo{volume}{533}},
  \bibinfo{pages}{118} (\bibinfo{year}{1986}).

\bibitem[{\citenamefont{{Dalmas de R\'eotier} and Yaouanc}(1997)}]{Dalmas97}
\bibinfo{author}{\bibfnamefont{P.}~\bibnamefont{{Dalmas de R\'eotier}}}
  \bibnamefont{and} \bibinfo{author}{\bibfnamefont{A.}~\bibnamefont{Yaouanc}},
  \bibinfo{journal}{J. Phys.: Condens. Matter} \textbf{\bibinfo{volume}{9}},
  \bibinfo{pages}{9113} (\bibinfo{year}{1997}).

\bibitem[{\citenamefont{{Dalmas de R\'eotier}
  et~al.}(2004)\citenamefont{{Dalmas de R\'eotier}, Gubbens, and
  Yaouanc}}]{Dalmas04}
\bibinfo{author}{\bibfnamefont{P.}~\bibnamefont{{Dalmas de R\'eotier}}},
  \bibinfo{author}{\bibfnamefont{P.~C.~M.} \bibnamefont{Gubbens}},
  \bibnamefont{and} \bibinfo{author}{\bibfnamefont{A.}~\bibnamefont{Yaouanc}},
  \bibinfo{journal}{J. Phys.: Condens. Matter} \textbf{\bibinfo{volume}{16}},
  \bibinfo{pages}{S4687} (\bibinfo{year}{2004}).

\bibitem[{\citenamefont{MacLaughlin et~al.}(2007)\citenamefont{MacLaughlin,
  Heffner, Nakatsuji, Nambu, Onuma, Maeno, Ishida, Bernal, and
  Shu}}]{MacLaughlin07}
\bibinfo{author}{\bibfnamefont{D.~E.} \bibnamefont{MacLaughlin}},
  \bibinfo{author}{\bibfnamefont{R.~H.} \bibnamefont{Heffner}},
  \bibinfo{author}{\bibfnamefont{S.}~\bibnamefont{Nakatsuji}},
  \bibinfo{author}{\bibfnamefont{Y.}~\bibnamefont{Nambu}},
  \bibinfo{author}{\bibfnamefont{K.}~\bibnamefont{Onuma}},
  \bibinfo{author}{\bibfnamefont{Y.}~\bibnamefont{Maeno}},
  \bibinfo{author}{\bibfnamefont{K.}~\bibnamefont{Ishida}},
  \bibinfo{author}{\bibfnamefont{O.~O.} \bibnamefont{Bernal}},
  \bibnamefont{and} \bibinfo{author}{\bibfnamefont{L.}~\bibnamefont{Shu}},
  \bibinfo{journal}{J. Mag. Mag. Mat.} \textbf{\bibinfo{volume}{310}},
  \bibinfo{pages}{1300} (\bibinfo{year}{2007}).

\bibitem[{\citenamefont{Overhauser}(1960)}]{Overhauser60}
\bibinfo{author}{\bibfnamefont{A.~W.} \bibnamefont{Overhauser}},
  \bibinfo{journal}{J. Phys. and Chem. Solids} \textbf{\bibinfo{volume}{13}},
  \bibinfo{pages}{71} (\bibinfo{year}{1960}).

\bibitem[{\citenamefont{{Dalmas de R\'eotier}
  et~al.}(2006)\citenamefont{{Dalmas de R\'eotier}, Yaouanc, Keller,
  Cervellino, Roessli, Baines, Forget, Vaju, Gubbens, Amato et~al.}}]{Dalmas06}
\bibinfo{author}{\bibfnamefont{P.}~\bibnamefont{{Dalmas de R\'eotier}}},
  \bibinfo{author}{\bibfnamefont{A.}~\bibnamefont{Yaouanc}},
  \bibinfo{author}{\bibfnamefont{L.}~\bibnamefont{Keller}},
  \bibinfo{author}{\bibfnamefont{A.}~\bibnamefont{Cervellino}},
  \bibinfo{author}{\bibfnamefont{B.}~\bibnamefont{Roessli}},
  \bibinfo{author}{\bibfnamefont{C.}~\bibnamefont{Baines}},
  \bibinfo{author}{\bibfnamefont{A.}~\bibnamefont{Forget}},
  \bibinfo{author}{\bibfnamefont{C.}~\bibnamefont{Vaju}},
  \bibinfo{author}{\bibfnamefont{P.~C.~M.} \bibnamefont{Gubbens}},
  \bibinfo{author}{\bibfnamefont{A.}~\bibnamefont{Amato}},
  \bibnamefont{et~al.}, \bibinfo{journal}{Phys. Rev. Lett.}
  \textbf{\bibinfo{volume}{96}}, \bibinfo{eid}{127202} (\bibinfo{year}{2006}).

\bibitem[{Yao()}]{Yaouanc_Dalmas_unpublished_1}
\bibinfo{note}{More specifically, through a combined used of the RPA
  approximation for the wavevector-dependent susceptibility and the
  fluctuation-dissipation and Kramers-Kroning theorems, an expression for the
  spin correlation function at zero frequency $\Lambda({\bf q},\omega=0)$ is
  derived. $\lambda_Z$ is proportional to a weighted sum over the Brillouin
  zone of $\Lambda({\bf q},\omega=0)$. Assuming that $\lambda_Z(T)$ is mainly
  driven by the correlation at a given wavevector (the magnetic structure
  propagation vector), the result given in the main text is found. The detailed
  derivation will be published elsewhere.}

\bibitem[{\citenamefont{Chaikin and Lubensky}(1995)}]{Chaikin95}
\bibinfo{author}{\bibfnamefont{P.~M.} \bibnamefont{Chaikin}} \bibnamefont{and}
  \bibinfo{author}{\bibfnamefont{T.~C.} \bibnamefont{Lubensky}},
  \emph{\bibinfo{title}{Principles of Condensed Matter Physics}}
  (\bibinfo{publisher}{Cambridge University Press},
  \bibinfo{address}{Cambridge}, \bibinfo{year}{1995}).

\bibitem[{\citenamefont{Nishiyama et~al.}(1984)\citenamefont{Nishiyama, Yagi,
  Ishida, Matsuzaki, Nagamine, and Yamazaki}}]{Nishiyama84}
\bibinfo{author}{\bibfnamefont{K.}~\bibnamefont{Nishiyama}},
  \bibinfo{author}{\bibfnamefont{E.}~\bibnamefont{Yagi}},
  \bibinfo{author}{\bibfnamefont{K.}~\bibnamefont{Ishida}},
  \bibinfo{author}{\bibfnamefont{T.}~\bibnamefont{Matsuzaki}},
  \bibinfo{author}{\bibfnamefont{K.}~\bibnamefont{Nagamine}}, \bibnamefont{and}
  \bibinfo{author}{\bibfnamefont{T.}~\bibnamefont{Yamazaki}},
  \bibinfo{journal}{Hyperfine Interactions} \textbf{\bibinfo{volume}{17-19}},
  \bibinfo{pages}{473} (\bibinfo{year}{1984}).

\bibitem[{\citenamefont{Herlach et~al.}(1986)\citenamefont{Herlach,
  K.~F\'urderer, and Schimmele}}]{Herlach86}
\bibinfo{author}{\bibfnamefont{D.}~\bibnamefont{Herlach}},
  \bibinfo{author}{\bibfnamefont{M.~F.} \bibnamefont{K.~F\'urderer}},
  \bibnamefont{and}
  \bibinfo{author}{\bibfnamefont{L.}~\bibnamefont{Schimmele}},
  \bibinfo{journal}{Hyperfine Interactions} \textbf{\bibinfo{volume}{31}},
  \bibinfo{pages}{287} (\bibinfo{year}{1986}).

\bibitem[{\citenamefont{Yaouanc et~al.}(1996)\citenamefont{Yaouanc, {Dalmas de
  R\'eotier}, Gubbens, Mulders, Kayzel, and Franse}}]{Yaouanc96}
\bibinfo{author}{\bibfnamefont{A.}~\bibnamefont{Yaouanc}},
  \bibinfo{author}{\bibfnamefont{P.}~\bibnamefont{{Dalmas de R\'eotier}}},
  \bibinfo{author}{\bibfnamefont{P.}~\bibnamefont{Gubbens}},
  \bibinfo{author}{\bibfnamefont{A.~M.} \bibnamefont{Mulders}},
  \bibinfo{author}{\bibfnamefont{F.~E.} \bibnamefont{Kayzel}},
  \bibnamefont{and} \bibinfo{author}{\bibfnamefont{J.~J.~M.}
  \bibnamefont{Franse}}, \bibinfo{journal}{Phys. Rev. B}
  \textbf{\bibinfo{volume}{53}}, \bibinfo{pages}{350} (\bibinfo{year}{1996}).

\bibitem[{\citenamefont{Phillips}(1996)}]{Phillips96}
\bibinfo{author}{\bibfnamefont{J.~C.} \bibnamefont{Phillips}},
  \bibinfo{journal}{Rep. Prog. Phys.} \textbf{\bibinfo{volume}{59}},
  \bibinfo{pages}{1133} (\bibinfo{year}{1996}).

\bibitem[{\citenamefont{Johnston}(2006)}]{Johnston06}
\bibinfo{author}{\bibfnamefont{D.~C.} \bibnamefont{Johnston}},
  \bibinfo{journal}{Phys. Rev. B} \textbf{\bibinfo{volume}{74}},
  \bibinfo{pages}{184430} (\bibinfo{year}{2006}).

\bibitem[{\citenamefont{McHenry et~al.}(1972)\citenamefont{McHenry,
  Silbernagel, and Wernick}}]{McHenry72}
\bibinfo{author}{\bibfnamefont{M.~R.} \bibnamefont{McHenry}},
  \bibinfo{author}{\bibfnamefont{B.~G.} \bibnamefont{Silbernagel}},
  \bibnamefont{and} \bibinfo{author}{\bibfnamefont{J.~H.}
  \bibnamefont{Wernick}}, \bibinfo{journal}{Phys. Rev. B}
  \textbf{\bibinfo{volume}{5}}, \bibinfo{pages}{2958} (\bibinfo{year}{1972}).

\bibitem[{\citenamefont{Uemura}(1980)}]{Uemura80}
\bibinfo{author}{\bibfnamefont{Y.~J.} \bibnamefont{Uemura}},
  \bibinfo{journal}{Solid State Commun.} \textbf{\bibinfo{volume}{36}},
  \bibinfo{pages}{369} (\bibinfo{year}{1980}).

\bibitem[{\citenamefont{Chapuis et~al.}(2007)\citenamefont{Chapuis, Yaouanc,
  {Dalmas de R\'eotier}, Pouget, Fouquet, Cervellino, and Forget}}]{Chapuis07}
\bibinfo{author}{\bibfnamefont{Y.}~\bibnamefont{Chapuis}},
  \bibinfo{author}{\bibfnamefont{A.}~\bibnamefont{Yaouanc}},
  \bibinfo{author}{\bibfnamefont{P.}~\bibnamefont{{Dalmas de R\'eotier}}},
  \bibinfo{author}{\bibfnamefont{S.}~\bibnamefont{Pouget}},
  \bibinfo{author}{\bibfnamefont{P.}~\bibnamefont{Fouquet}},
  \bibinfo{author}{\bibfnamefont{A.}~\bibnamefont{Cervellino}},
  \bibnamefont{and} \bibinfo{author}{\bibfnamefont{A.}~\bibnamefont{Forget}},
  \bibinfo{journal}{J. Phys.: Condens. Matter} \textbf{\bibinfo{volume}{19}},
  \bibinfo{pages}{446206} (\bibinfo{year}{2007}).

\bibitem[{\citenamefont{Rule et~al.}(2007)\citenamefont{Rule, Ehlers, Stewart,
  Cornelius, Deen, Qiu, Wiebe, Janik, Zhou, Antonio et~al.}}]{Rule07}
\bibinfo{author}{\bibfnamefont{K.~C.} \bibnamefont{Rule}},
  \bibinfo{author}{\bibfnamefont{G.}~\bibnamefont{Ehlers}},
  \bibinfo{author}{\bibfnamefont{J.~R.} \bibnamefont{Stewart}},
  \bibinfo{author}{\bibfnamefont{A.~L.} \bibnamefont{Cornelius}},
  \bibinfo{author}{\bibfnamefont{P.~P.} \bibnamefont{Deen}},
  \bibinfo{author}{\bibfnamefont{Y.}~\bibnamefont{Qiu}},
  \bibinfo{author}{\bibfnamefont{C.~R.} \bibnamefont{Wiebe}},
  \bibinfo{author}{\bibfnamefont{J.~A.} \bibnamefont{Janik}},
  \bibinfo{author}{\bibfnamefont{H.~D.} \bibnamefont{Zhou}},
  \bibinfo{author}{\bibfnamefont{D.}~\bibnamefont{Antonio}},
  \bibnamefont{et~al.}, \bibinfo{journal}{Phys. Rev. B}
  \textbf{\bibinfo{volume}{76}}, \bibinfo{eid}{212405} (\bibinfo{year}{2007}).

\bibitem[{\citenamefont{Mazin}(2007)}]{Mazin07}
\bibinfo{author}{\bibfnamefont{I.~I.} \bibnamefont{Mazin}},
  \bibinfo{journal}{Phys. Rev. B} \textbf{\bibinfo{volume}{76}},
  \bibinfo{pages}{140406} (\bibinfo{year}{2007}).

\bibitem[{\citenamefont{Takubo et~al.}(2007)\citenamefont{Takubo, Mizokawa,
  Son, Nambu, Nakatsuji, and Maeno}}]{Takubo07}
\bibinfo{author}{\bibfnamefont{K.}~\bibnamefont{Takubo}},
  \bibinfo{author}{\bibfnamefont{T.}~\bibnamefont{Mizokawa}},
  \bibinfo{author}{\bibfnamefont{J.-Y.} \bibnamefont{Son}},
  \bibinfo{author}{\bibfnamefont{Y.}~\bibnamefont{Nambu}},
  \bibinfo{author}{\bibfnamefont{S.}~\bibnamefont{Nakatsuji}},
  \bibnamefont{and} \bibinfo{author}{\bibfnamefont{Y.}~\bibnamefont{Maeno}},
  \bibinfo{journal}{Phys. Rev. Lett.} \textbf{\bibinfo{volume}{99}},
  \bibinfo{pages}{037203} (\bibinfo{year}{2007}).

\bibitem[{\citenamefont{Chandra and Coleman}(1991)}]{Chandra91}
\bibinfo{author}{\bibfnamefont{P.}~\bibnamefont{Chandra}} \bibnamefont{and}
  \bibinfo{author}{\bibfnamefont{P.}~\bibnamefont{Coleman}},
  \bibinfo{journal}{Phys. Rev. Lett.} \textbf{\bibinfo{volume}{66}},
  \bibinfo{pages}{100} (\bibinfo{year}{1991}).

\bibitem[{\citenamefont{Tsunetsugu and Arikawa}(2006)}]{Tsunetsugu06}
\bibinfo{author}{\bibfnamefont{H.}~\bibnamefont{Tsunetsugu}} \bibnamefont{and}
  \bibinfo{author}{\bibfnamefont{M.}~\bibnamefont{Arikawa}},
  \bibinfo{journal}{J. Phys. Soc. Japan} \textbf{\bibinfo{volume}{75}},
  \bibinfo{pages}{083701} (\bibinfo{year}{2006}).

\bibitem[{\citenamefont{L\"auchli et~al.}(2006)\citenamefont{L\"auchli, Mila,
  and Penc}}]{Lauchli06}
\bibinfo{author}{\bibfnamefont{A.}~\bibnamefont{L\"auchli}},
  \bibinfo{author}{\bibfnamefont{F.}~\bibnamefont{Mila}}, \bibnamefont{and}
  \bibinfo{author}{\bibfnamefont{K.}~\bibnamefont{Penc}},
  \bibinfo{journal}{Phys. Rev. Lett.} \textbf{\bibinfo{volume}{97}},
  \bibinfo{pages}{087205} (\bibinfo{year}{2006}).

\bibitem[{\citenamefont{Bhattacharjee et~al.}(2006)\citenamefont{Bhattacharjee,
  Shenoy, and Senthil}}]{Bhattacharjee06}
\bibinfo{author}{\bibfnamefont{S.}~\bibnamefont{Bhattacharjee}},
  \bibinfo{author}{\bibfnamefont{V.~B.} \bibnamefont{Shenoy}},
  \bibnamefont{and} \bibinfo{author}{\bibfnamefont{T.}~\bibnamefont{Senthil}},
  \bibinfo{journal}{Phys. Rev. B} \textbf{\bibinfo{volume}{74}},
  \bibinfo{pages}{092406} (\bibinfo{year}{2006}).

\bibitem[{\citenamefont{Li et~al.}(2007{\natexlab{a}})\citenamefont{Li, Zhang,
  and Shen}}]{Li07a}
\bibinfo{author}{\bibfnamefont{P.}~\bibnamefont{Li}},
  \bibinfo{author}{\bibfnamefont{G.~M.} \bibnamefont{Zhang}}, \bibnamefont{and}
  \bibinfo{author}{\bibfnamefont{S.~Q.} \bibnamefont{Shen}},
  \bibinfo{journal}{Phys. Rev. B} \textbf{\bibinfo{volume}{75}},
  \bibinfo{pages}{104420} (\bibinfo{year}{2007}{\natexlab{a}}).

\bibitem[{\citenamefont{Kawamura and Miyashita}(1984)}]{Kawamura84}
\bibinfo{author}{\bibfnamefont{H.}~\bibnamefont{Kawamura}} \bibnamefont{and}
  \bibinfo{author}{\bibfnamefont{S.}~\bibnamefont{Miyashita}},
  \bibinfo{journal}{J. Phys. Soc. Japan} \textbf{\bibinfo{volume}{53}},
  \bibinfo{pages}{4138} (\bibinfo{year}{1984}).

\bibitem[{\citenamefont{Kawamura and Yamamoto}(2007)}]{Kawamura07}
\bibinfo{author}{\bibfnamefont{H.}~\bibnamefont{Kawamura}} \bibnamefont{and}
  \bibinfo{author}{\bibfnamefont{A.}~\bibnamefont{Yamamoto}},
  \bibinfo{journal}{J. Phys. Soc. Japan} \textbf{\bibinfo{volume}{76}},
  \bibinfo{pages}{073704} (\bibinfo{year}{2007}).

\bibitem[{\citenamefont{Fujimoto}(2006)}]{Fujimoto06}
\bibinfo{author}{\bibfnamefont{S.}~\bibnamefont{Fujimoto}},
  \bibinfo{journal}{Phys. Rev. B} \textbf{\bibinfo{volume}{73}},
  \bibinfo{pages}{184401} (\bibinfo{year}{2006}).

\bibitem[{\citenamefont{Azaria et~al.}(1992)\citenamefont{Azaria, Delamotte,
  and Mouhanna}}]{Azaria92}
\bibinfo{author}{\bibfnamefont{P.}~\bibnamefont{Azaria}},
  \bibinfo{author}{\bibfnamefont{B.}~\bibnamefont{Delamotte}},
  \bibnamefont{and} \bibinfo{author}{\bibfnamefont{D.}~\bibnamefont{Mouhanna}},
  \bibinfo{journal}{Phys. Rev. Lett.} \textbf{\bibinfo{volume}{68}},
  \bibinfo{pages}{1762} (\bibinfo{year}{1992}).

\bibitem[{\citenamefont{Southern and Young}(1993)}]{Southern93}
\bibinfo{author}{\bibfnamefont{B.~W.} \bibnamefont{Southern}} \bibnamefont{and}
  \bibinfo{author}{\bibfnamefont{A.~P.} \bibnamefont{Young}},
  \bibinfo{journal}{Phys. Rev. B} \textbf{\bibinfo{volume}{48}},
  \bibinfo{pages}{R13170} (\bibinfo{year}{1993}).

\bibitem[{\citenamefont{Wintel et~al.}(1995)\citenamefont{Wintel, Everts, and
  Apel}}]{Wintel95}
\bibinfo{author}{\bibfnamefont{M.}~\bibnamefont{Wintel}},
  \bibinfo{author}{\bibfnamefont{H.~U.} \bibnamefont{Everts}},
  \bibnamefont{and} \bibinfo{author}{\bibfnamefont{W.}~\bibnamefont{Apel}},
  \bibinfo{journal}{Phys. Rev. B} \textbf{\bibinfo{volume}{52}},
  \bibinfo{pages}{13480} (\bibinfo{year}{1995}).

\bibitem[{\citenamefont{Li et~al.}(2007{\natexlab{b}})\citenamefont{Li, Su, and
  Shen}}]{Li07b}
\bibinfo{author}{\bibfnamefont{P.}~\bibnamefont{Li}},
  \bibinfo{author}{\bibfnamefont{H.}~\bibnamefont{Su}}, \bibnamefont{and}
  \bibinfo{author}{\bibfnamefont{S.~Q.} \bibnamefont{Shen}},
  \bibinfo{journal}{Phys. Rev. B} \textbf{\bibinfo{volume}{76}},
  \bibinfo{pages}{174406} (\bibinfo{year}{2007}{\natexlab{b}}).

\bibitem[{\citenamefont{Yaouanc et~al.}(2005)\citenamefont{Yaouanc, {P. Dalmas
  de R\'eotier}, Glazkov, Marin, Bonville, Hodges, Gubbens, Sakarya, and
  Baines}}]{Yaouanc05a}
\bibinfo{author}{\bibfnamefont{A.}~\bibnamefont{Yaouanc}},
  \bibinfo{author}{\bibnamefont{{P. Dalmas de R\'eotier}}},
  \bibinfo{author}{\bibfnamefont{V.}~\bibnamefont{Glazkov}},
  \bibinfo{author}{\bibfnamefont{C.}~\bibnamefont{Marin}},
  \bibinfo{author}{\bibfnamefont{P.}~\bibnamefont{Bonville}},
  \bibinfo{author}{\bibfnamefont{J.~A.} \bibnamefont{Hodges}},
  \bibinfo{author}{\bibfnamefont{P.~C.~M.} \bibnamefont{Gubbens}},
  \bibinfo{author}{\bibfnamefont{S.}~\bibnamefont{Sakarya}}, \bibnamefont{and}
  \bibinfo{author}{\bibfnamefont{C.}~\bibnamefont{Baines}},
  \bibinfo{journal}{Phys. Rev. Lett.} \textbf{\bibinfo{volume}{95}},
  \bibinfo{eid}{047203} (\bibinfo{year}{2005}).

\bibitem[{\citenamefont{Takeya et~al.}(2007)\citenamefont{Takeya, Ishida,
  Kitagawa, Ihara, Onuma, Maeno, Nambu, Nakatsuji, MacLaughlin, Koda
  et~al.}}]{Takeya08}
\bibinfo{author}{\bibfnamefont{H.}~\bibnamefont{Takeya}},
  \bibinfo{author}{\bibfnamefont{K.}~\bibnamefont{Ishida}},
  \bibinfo{author}{\bibfnamefont{K.}~\bibnamefont{Kitagawa}},
  \bibinfo{author}{\bibfnamefont{Y.}~\bibnamefont{Ihara}},
  \bibinfo{author}{\bibfnamefont{K.}~\bibnamefont{Onuma}},
  \bibinfo{author}{\bibfnamefont{Y.}~\bibnamefont{Maeno}},
  \bibinfo{author}{\bibfnamefont{Y.}~\bibnamefont{Nambu}},
  \bibinfo{author}{\bibfnamefont{S.}~\bibnamefont{Nakatsuji}},
  \bibinfo{author}{\bibfnamefont{D.~E.} \bibnamefont{MacLaughlin}},
  \bibinfo{author}{\bibfnamefont{K.}~\bibnamefont{Koda}}, \bibnamefont{et~al.},
  \bibinfo{journal}{arXiv:0801.0190}  (\bibinfo{year}{2007}).

\end{thebibliography}

\end{document}